\documentclass[preprint,showpacs,showkeys,preprintnumbers,aps]{revtex4}

\usepackage{graphicx}
\usepackage{float}
\usepackage{amsmath}
\usepackage{braket}

\begin{document}

\title{Topological insulators and thermoelectric materials}


\author{Lukas~M\"uchler}
\affiliation{Institut f\"ur Anorganische und Analytische Chemie,
Johannes Gutenberg - Universit\"at, 55099 Mainz, Germany.}
\author{Frederick Casper}
\affiliation{Institut f\"ur Anorganische und Analytische Chemie,
Johannes Gutenberg - Universit\"at, 55099 Mainz, Germany.}
\author{Binghai~Yan}
\affiliation{Institut f\"ur Anorganische und Analytische Chemie,
Johannes Gutenberg - Universit\"at, 55099 Mainz, Germany.}
\author{Stanislav~Chadov}
\affiliation{Max-Planck-Institute f\"ur Chemische Physik fester Stoffe, N\"othnitzer Str.~40, 01187 Dresden, Germany.}
\author{Claudia~Felser}
\affiliation{Max-Planck-Institute f\"ur Chemische Physik fester Stoffe, N\"othnitzer Str.~40, 01187 Dresden, Germany.}
\email{felser@cpfs.mpg.de}


\begin{abstract}
Topological insulators (TIs) are a new quantum state of matter which have gapless surface states inside the bulk energy gap~\cite{QZh+10,HKa+10,MOO10,qi2011rmp}. Starting with the discovery of two dimensional TIs, the HgTe-based quantum wells~\cite{BHZ06,KWB+07}, many new topological materials have been theoretically predicted and experimentally observed. Currently known TI materials can possibly be classified into two families~\cite{yan2012topological}, the HgTe family and the Bi$_2$Se$_3$ family. The signatures found in the electronic structure of a TI also cause these materials to be excellent thermoelectric materials~\cite{kong2011,MZC+12,HRR09}. 
On the other hand, excellent thermoelectric materials can be also topologically trivial. Here we present a short introduction to topological insulators and thermoelectrics, and give examples of compound classes were both good thermoelectric properties and topological insulators can be found.
\end{abstract}

\maketitle   

\section{Introduction}

\begin{figure}[H]%
  \includegraphics[width=0.7\linewidth]{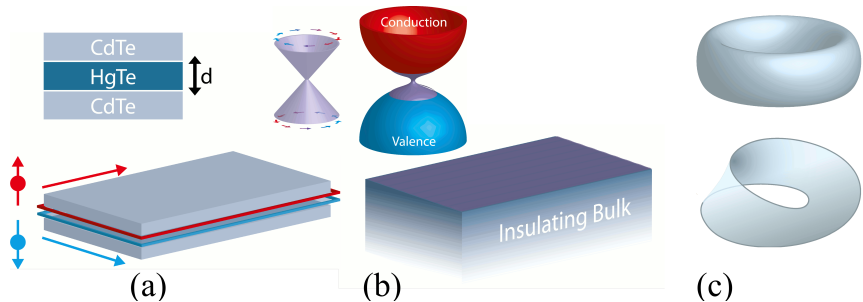}%
  \caption[]{%
    Schematics of a 2D TI (HgTe/CdTe quantum wells) with gapless edge states (a) and a 3D TI with
gapless surface states resembling a Dirac cone (b). The spin direction is labeled for the topological
states. (c) An example of two topological distinct objects: a regular strip and a
M\"obius strip. }
\label{Topo}
\end{figure}

Topological Insulators (TIs) have generated a great
interest in the fields of condensed matter physics, chemistry and materials
science ~\cite{QZh+10,HKa+10,MOO10,qi2011rmp,yan2012topological,kong2011,MZC+12}. 
Generally, compounds can be classified into metals and insulators according to their band structures. However, the TI is an insulator in the bulk, while it 
conducts on the surface like a metal. TIs has 2D (also called quantum spin Hall (QSH) state) and 3D versions.
They have a full energy gap in the bulk, but host topologically protected gapless edge (2D) or surface (3D) states (see Fig.~\ref{Topo}).

A TI can easily be identified by a few simple rules: the
presence of large spin orbit coupling (SOC), an odd number of band inversions (BI)
between the conduction and the valence band by increasing the average nuclear
charge, and a sign change of the symmetry of the molecular
orbitals~\cite{MZC+12}. Many TIs were known as excellent thermoelectric (TE) materials, for TI and TE compounds
usually favor the same material features, such as heavy elements and the smaller energy gap. 
Thus TIs maybe good TE candidates and vice
versa. Here we present a short overview on to TIs and give
examples of compound classes were both TIs and good TE properties can be found. 

\section{Thermoelectrics and topological insulators}

\subsection{Thermoelectrics}

\begin{figure}[H]%
  \includegraphics[width=0.7\linewidth]{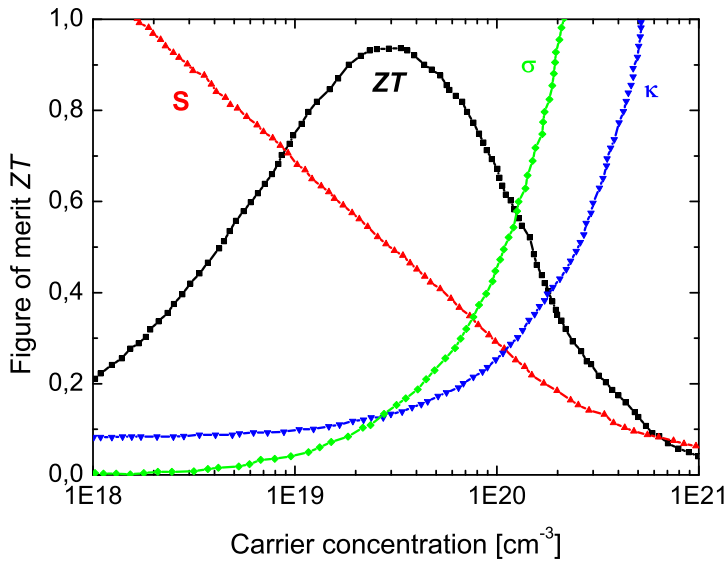}
  \caption[]{%
    Correlation of the  electrical conductivity $\sigma$, Seebeck coefficient \textit{S}, thermal conductivity $\kappa$ and carrier concentration. The optimum figure of merit is reached via a compromise of the three parameters. (data taken from~\cite{snyder11})}
    \label{Thermo}
\end{figure}

In principle, in thermoelectric materials a temperature gradient creates an electric potential, and thus these materials could be used as TE converters for power generation. TE converter can be used for the conversion of a part of the low-grade waste heat generated by engines, industrial furnaces, gas pipes, etc. to electricity. The performance of the TE materials is characterized by the figure of merit $ZT$:
\begin{equation}
\label{eq1}
ZT = \frac{ S^2 \sigma}{\kappa}T,
\end{equation}
where $\sigma$ is the electrical conductivity, \textit{S} the Seebeck coefficient and $\kappa$ the thermal conductivity. A larger value of $ZT$ means a higher energy conversion efficiency, which achieves its thermodynamic limit of the Carnot efficiency when $ZT$ is infinite. To reach a high figure of merit, a large Seebeck coefficient, a high electrical conductivity $\sigma$, and a low thermal conductivity $\kappa$ are required. Unfortunately, these parameters are connected: a high Seebeck coefficient prefers a low carrier concentration, which leads to a low electrical conductivity; the high electrical conductivity comes together with high thermal conductivity ($\kappa_e$). 
As shown in figure~\ref{Thermo}, the maximum $ZT$ of bulk Bi$_2$Te$_3$ lies in an optimal region where the carrier density is
between a semiconductor and a metal. High performance TE materials are are typically heavily doped semiconductors to give metallic bulk conductivity.

To increase the electrical conductivity, narrow band-gap materials with high mobilities are a natural choice. In these materials, the energy gap can be tuned (e.g. by chemical compositions) to optimize the TE performance. To reduce the thermal conductivity, elements with large atomic mass are favored. Besides changing these intrinsic properties, other approaches are also employed to reduce the thermal conductivity~\cite{GRA11}. For examples, introducing grain boundaries and forming alloys can dramatically enhance the phonon scattering.

\subsection{Topological Insulators}
TIs can occur in 2- and 3-dimensional materials (Fig.~\ref{Topo}\,a,b).
The 2D TI has wire-like metallic edge states, in which counter-propagating electrons have opposite spin.
A 3D TI has surface states around the entire material, usually forming a single or an odd number of Dirac cones.
In the bulk however, TIs are usually narrow gap semiconductors.
It is possible to distinct regular semiconductors such as Si, which do not possess such edge or surface states, from TIs by the topology of their band structure. The condition of a material to be a TI is an odd number of BIs in the bulk due to relativistic effects, especially SOC. 
Just as a M\"obius strip is different from a regular strip of paper in its topology (Fig.~\ref{Topo}\,c), TIs are distinct from regular semiconductors and insulators due to the topology of their band structures\cite{MBa+07,Roy+09,KMe+05,KMe+05b,FKa+06,QHZ+08}. 
Important for the discussion above is the conservation of time reversal symmetry (TRS).
Here TRS means that there are no external magnetic fields, magnetic moments and magnetic impurities in the system. 
The interface state of a TI is inherently robust against nonmagnetic impurities and surface disorders, distinguished from normal surface states such as dangling bonds. 
In a TI, the forward- and backward-moving topological edge states are a pair of time reversal conjugates, i.e. $\Theta \ket{\uparrow, k} = \ket{\downarrow, -k}$ with $\Theta$ beeing the time reversal operator, and therefore have opposite spin orientation.
In this way, an electron cannot be scattered from a forward-moving state into a backward-moving one without reversing its spin. 
Hence backscattering is suppressed under TRS due to this spin-momentum locking.

\subsection{Electronic structure of topological insulators}
\begin{figure}[H]%
  \includegraphics[width=0.7\linewidth]{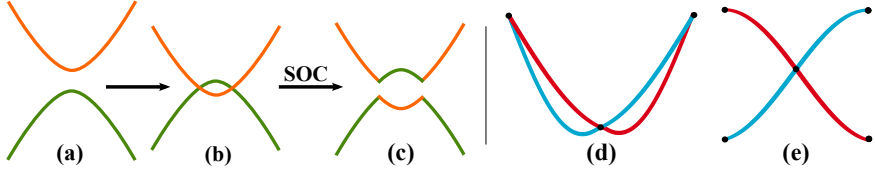}%
  \caption[]{%
    Schematic band structure of a trivial semiconductor (a). Due
to strong relativistic effects, the conduction and valence bands cross each other and get inverted (b).
Spin-orbit coupling opens a gap due to band anti-crossing (c).
Difference between the surface states of a Rashba-split system (d) and
a TI (e) . Each color represents a different in-plane spin
component and black dots indicate time-reversal invariant momenta which show a
Kramer's degeneracy}
    \label{Bands}
\end{figure}

The necessary condition for a compound to be a TI is an odd number of band
inversions (BIs) between conduction and valence bands ~\cite{MZC+12}. 
This BI is due to relativistic effects such as SOC, which can be enhanced by
substituting light elements with heavier elements from the same group.
As illustrated in Fig.~\ref{Bands}, the conduction and valence bands get inverted and cross each other when SOC is strong enough.
A new energy gap emerges at the crossing points due to SOC, called band anti-crossing. 
If this type of BI happens odd times in the whole Brillouin zone, a TI forms.
It is known that strong SOC leads to spin split states on the surface due to the Rashba-effect. (see
Fig.~\ref{Bands} (d) ). However, the topological surface states are different in topology from regular Rashba-split states. 
The former connects the conduction and valence bands, and can not be gapped away. In contrast,
the latter is only related to the conduction or valence bands, and can be deformed smoothly into the bulk bands, equivalent to 
a fully gapped band structure.

\begin{figure}[H]%
  \includegraphics[width=0.7\linewidth]{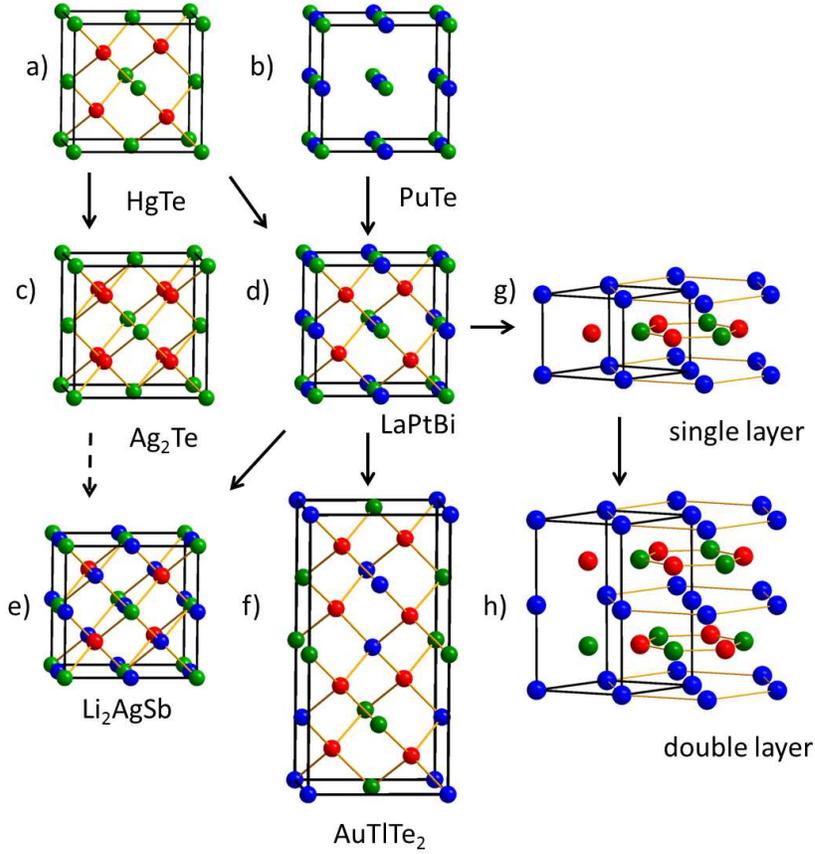}%
  \caption[]{%
    Crystal structures of topological compounds based on the cubic structure. a) HgTe and b) PuTe are the parent compounds with ZnS and NaCl structure, respectively; occupation of the tetrahedral vacancies of the ZnS structure leads to a family with an inverse CaF$_2$ structure (low-temperature $\alpha$-phase of Ag$_2$Te) c); filling up the octahedral vacancies in the ZnS structure or the tetrahedral vacancies ind the ZnS structure leads to the family of Heuslers (LaPtBi) d); by adding more "stuffing" atoms to a Heusler, the so-called "inverse'' Heusler structure is obtained (Li$_2$AgSb) e); doubling the cubic Heusler unit cell leads to a chalcopyrite structure (AuTlTe$_2$)(f).}
    \label{CubicStructures}
\end{figure}

\section{The HgTe family of compounds } 
\subsection{ HgTe}
In 2006, Bernevig, Hughes and Zhang ~\cite{BHZ06} predicted HgTe quantum wells
(QWs) to be 2D topological insulators.
A QW consists of a thin HgTe layer sandwiched between two conventional
insulating layers of CdTe (Fig.~\ref{Topo}\,a). 
The low energy bands near the Fermi level of HgTe and CdTe are close to the $\Gamma$ point in the Brillouin zone,
as shown in Fig.\ref{zb-heusler_fingerprint}. Like the
common semiconductor GaAs, CdTe has a band ordering with $s$-orbital $\Gamma_6$ band as the conduction band minimum (CBM) and 
$p$-orbital $\Gamma_8$ as the
valence band maximum (VBM). The CBM and VBM are separated by a large energy gap $\sim 1.6$ eV.
Compared to Cd, Hg is a heavier element and has stronger relativistic effects. As a result of both
the Darwin term and the SOC, HgTe exhibits an ``inverted'' band ordering: The $p$ band is shifted above
$s$ band. Due to cubic symmetry these two $p$ bands are degenerate at the $\Gamma$ point.
Therefore, HgTe is a zero-gap semiconductor with an inverted band ordering. 

HgTe can open an energy gap when the cubic symmetry is broken by strain or by lowering the dimensionality.
The conventional semiconductor CdTe serves as an ideal barrier layer for QWs. When the QW thickness $d_{QW}$ is smaller 
than a critical value $d_c$, the QW should behave similarly to CdTe and has normal band ordering.
On the other hand, when $d_{QW} > d_c $, the QW should behave more
like HgTe with inverted bands, forming a 2D TI.
Soon after the theoretical prediction, the 2D TI was observed experimentally in
HgTe quantum wells~\cite{KWB+07}.
Both HgTe and CdTe crystallize in the zinc-blende structure (Fig.~\ref{CubicStructures}a), which is a relative
of the diamond structure. 
On the other hand, simple rocksalt compounds formed by Pu and Am (Fig.~\ref{CubicStructures}b) lie on the boundary between metals and insulators, where the 5f electrons change character from itinerant to localized. In comparison to standard TIs such as Bi$_2$Te$_3$ and HgTe, the actinide compounds have stronger three-dimensional ionic bonds. A single Dirac cone at $\Gamma$ point of e.g. AmN shows up in the energy gap, which demonstrates the nontrivial nature of the materials~\cite{ZZW12}.

\subsection{Heusler compounds}
\begin{figure}[ht]%
  \includegraphics[width=0.7\linewidth]{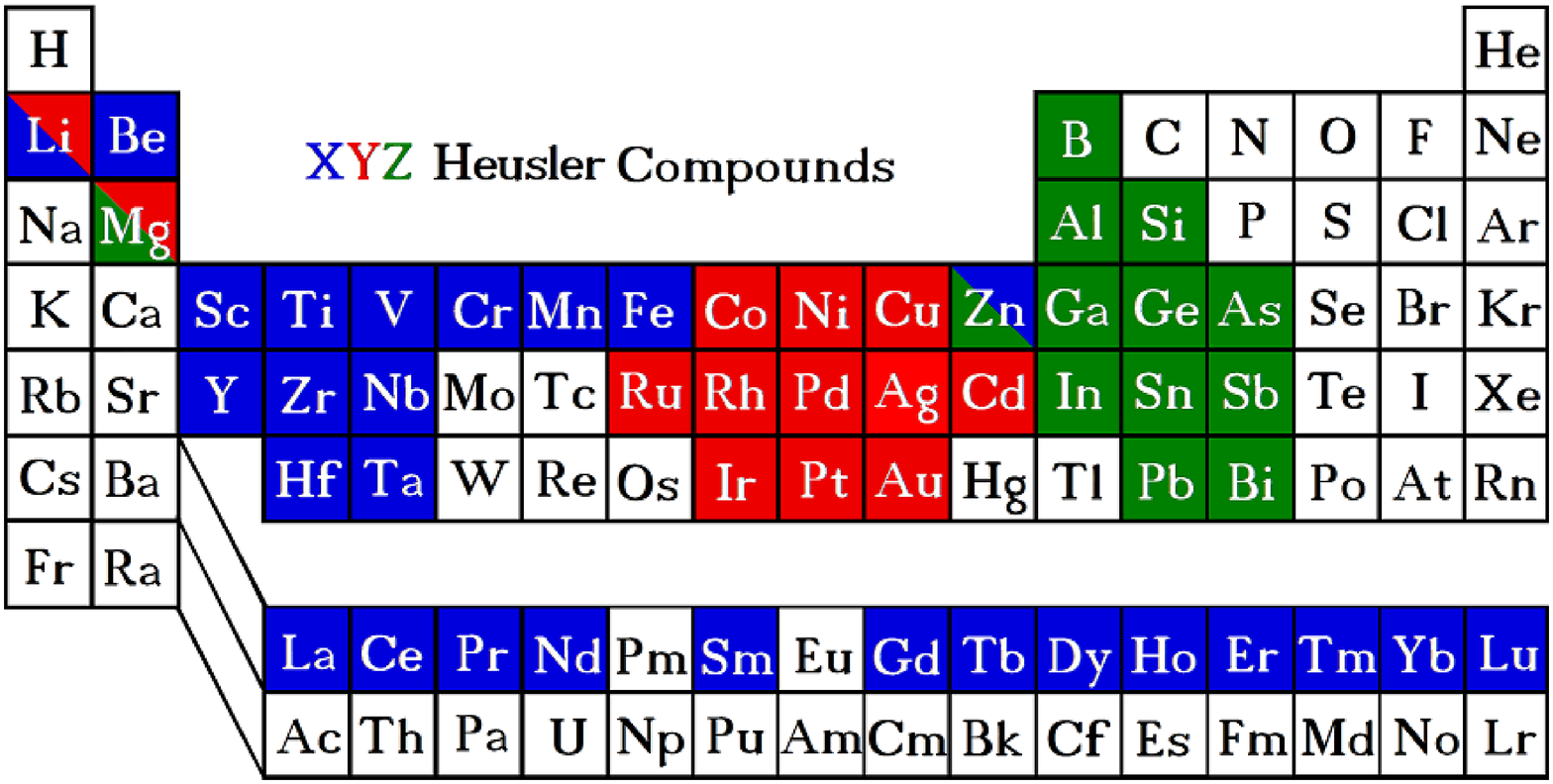}%
  \caption[]{%
    Possible element distribution of cubic \textit{XYZ} half-Heusler compounds}
    \label{PeriodictableHH}
\end{figure}

As analogs of HgTe, Heusler compounds are one of the most versatile class of compounds~\cite{GFP+11}. They are named after their discoverer Fritz Heusler~\cite{HSH+03,Heu+03}, and up to now this class of materials includes more than 1500 different compounds. Semiconducting behavior is mainly found in the subclass of the half-Heusler (\textit{XYZ}) compounds. A half-Heusler material \textit{XYZ} is a compound consisting of a covalent and an ionic part: The \textit{X}  and  \textit{Y} atoms  have a  distinct cationic  character,  whereas \textit{Z} can be seen as the anionic counterpart. The most electropositive element (usually a main group element, a transition metal or a rare earth element) is placed at the beginning of the formula. The most electronegative element is a main group element from the second half of the periodic table. Figure~\ref{PeriodictableHH} shows the element distribution of \textit{X}, \textit{Y} and \textit{Z} within the periodic table.

\begin{figure}[ht]%
  \includegraphics[width=0.7\linewidth]{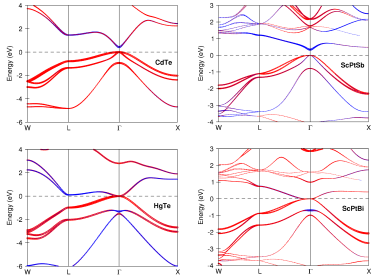}%
  \caption[]{%
    The bandstructures of binary zinc-blende and ternary Heusler
  semiconductors reveal clear fingerprints: both CdTe and
ScPtSb exhibit a direct gap at the $\Gamma$-point between the
conduction $s$-like (blue) and the valence $p$-like (red) bands of $\Gamma_6$ and
$\Gamma_8$ symmetries, respectively. On the other hand, the
band structures of both heavier compounds, HgTe and ScPtBi,
exhibit the band inversion caused by the strong spin-orbit coupling: $\Gamma_6$ (blue) is now situated below
$\Gamma_8$ (red) which remains at the Fermi energy (horizontal dashed line).}
    \label{zb-heusler_fingerprint}
\end{figure}

Generally, the half-Heusler phases crystallize in a non-centrosymmetric structure corresponding to the space group F$\overline{4}$3m (No.~216). Within the lattice, the atoms on Wyckoff positions $4a$ (0, 0, 0) and $4b$ ($\frac{1}{2}, \frac{1}{2}, \frac{1}{2}$) form the ionic NaCl-type substructure, while the atoms on $4a$ and $4c$ ($\frac{1}{4}, \frac{1}{4}, \frac{1}{4}$) build the covalent ZnS-type one (Fig.~\ref{CubicStructures}d). The properties of half-Heusler compounds can be predicted just by counting the number of their valence electrons~\cite{FFB07}, and the band gaps can be tuned from zero up to 4~eV by simply changing their chemical composition. Thus they attracted great interest in the field of thermoelectrics and topological insulators. Compounds with eight valence electrons per formula unit are closely related to classical semiconductors, such as silicon and GaAs. Within this class of materials, 3 subgroups have to be differentiated: The Nowotny-Juza phases A$^I$B$^{II}$C$^V$ with A$^I$ = Li, Cu, Ag, B$^{II}$ = Be, Mg, Zn, Cd, and C$^V$ = N, P, As
Sb, Bi, which are well known wide band gap semiconductors~\cite{JHu+48,NBa+50,KBe+06}, and the nameless A$^I$B$^{III}$C$^{IV}$ (for instance LiAlSi~\cite{SJD+03} and LiGaSi~\cite{NHo+60}) and the A$^{II}$B$^{II}$C$^{IV}$ phases e.\,g. Mg$_2$Si~\cite{Mar+72}). Within the 8 valence electron compounds, the gap size is larger for compounds with a large Pauli electronegativity difference of the \textit{Y} and \textit{Z} species~\cite{KFS+06}.  However, the design of unconventional semiconductors based on 18 valence electrons for half-Heusler compounds is also possible. These materials contain transition metal elements with almost completely filled $d$-electron shell which is added to the valence electron count, again leading to a closed-shell configuration and semiconducting properties. 
The broad class of Heusler systems was systematically investigated and many new ternary systems with non-trivial band structures were
suggested~\cite{CQK+10,LWX+10}. As half-Heusler compounds with 8- and 18- valence electrons are mainly semiconductors they are electronically related to the binary semiconductors (see Fig.~\ref{zb-heusler_fingerprint}).

Due to such similarity, the mechanisms responsible for the driving the
system into a certain topological class, are the same for the
zinc-blende as well as for the C1$_{b}$ Heusler compounds. 
The band inversion at the $\Gamma$ point, which drives these materials into trivial or
non-trivial topological class, solely depends on the interplay of two competing
mechanisms. First is the orbital $sp$-hybridization, which opens a real band gap between
$s$-like state of $\Gamma_6$-symmetry shifted into the unoccupied region
and the $p$-state containing $\Gamma_8$-symmetric representation, which
is partially occupied and thus degenerated at the Fermi energy. The
second mechanism is the spin-orbit coupling, which splits the
$p$-shell of the main-group element (e.\,g. Se, Te, Bi) into the higher $p_{3/2}$ ($\Gamma_8$-symmetric)
and lower-energy $p_{1/2}$ bands. If the spin-orbit coupling is strong enough, the $s$-like band pops 
inside this energy gap and gets occupied. This causes the desired band inversion. 

\begin{figure*}[htb]
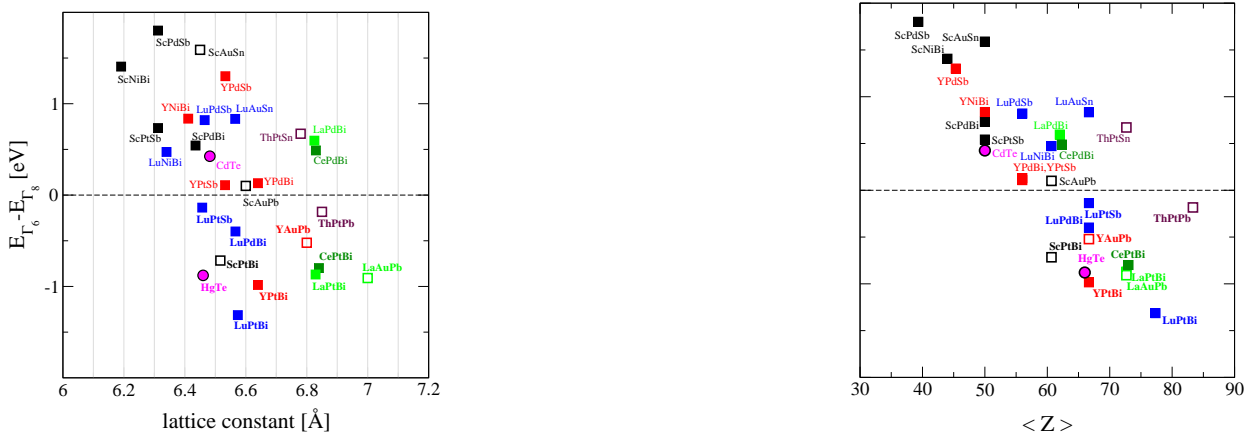
%
\includegraphics[width=.35\textwidth]{pssr.201206411_Fig8a.eps}\hfill
\includegraphics[width=.32\textwidth]{pssr.201206411_Fig8b.eps}\hfill
\caption{%
The energy difference between $\Gamma_6$
and $\Gamma_8$ bands calculated as a function of lattice constant (right)
and the mean nuclear number $\langle Z\rangle$ (left)~\cite{CQK+10}. Each subgroup (\textit{Ln}NiSb, \textit{Ln}PdSb, \textit{Ln}PdBi, \textit{Ln}PtBi,
\textit{Ln}AuSn, \textit{Ln}AuPb and ThPt\textit{Z}) is marked by a certain color. Different
shapes of markers (squares, circle etc.) are assigned to a
different stuffing elements (Sc, Y, La, Lu, Th). 
Compounds with ${E_{\Gamma_6}-E_{\Gamma_8}>0}$ 
 are topologically trivial, whereas those with
 ${E_{\Gamma_6}-E_{\Gamma_8}<0}$ are the non-trivial candidates.}
\label{diagram}
\end{figure*}

The diversity of Heusler materials opens wide possibilities to tune
the band inversion by choosing compounds with appropriate
hybridization strength (by lattice parameter) and the magnitude of
spin-orbit coupling (by the nuclear charge number $\langle Z\rangle =
1/N\sum_{i=1}^N Z_i$, where $N$ is the number of atoms in the unit cell) as shown in Fig.~\ref{diagram}. 

The great advantage of ternary compounds is their multifunctionality.
Many of these ternary zero-gap semiconductors (\textit{Ln}AuPb,
\textit{Ln}PdBi, \textit{Ln}PtSb and \textit{Ln}PtBi) contain the rare-earth element \textit{Ln} which
can be used alternatively for managing the additional properties
ranging from superconductivity (e.g. LaPtBi~\cite{GMH+08})
to    antiferromagnetism (e.g. GdPtBi~\cite{CTB+91})
and heavy-fermion  behavior (e.g. YbPtBi~\cite{FCB+91}). 

As shown in Fig.~\ref{CubicStructures}, there are more newly identified TI systems related to the HgTe structure: inverse Heusler compounds like Li$_2$AgSb (Fig.~\ref{CubicStructures}e) with all tetrahedral and octahedral sides filled in the ccp structure of the main group metal (Sb, Bi, or Te)~\cite{LWX10}. Another variant is Ag$_2$Te in the high-temperature cubic phase~\cite{ZYF11}, corresponding to an inverse CaF$_2$ structure (Fig.~\ref{CubicStructures}c) with fully occupied tetrahedral positions. A second low-symmetry structure family are the chalcopyrites. Their structure can be viewed as a doubled zinc-blende cell with a trivalent In or Ga and a divalent Zn substituted by monovalent Cu (Fig.~\ref{CubicStructures}f). The tetragonal symmetry breaks the x,y,z degeneracy thus opens a finite band gap~\cite{FXD11}.

Similarly to the HgTe, all topologically-nontrivial semiconducting
Heusler systems with C1$_b$ structure are zero-gap semiconductors
(often called semimetals), rather than real insulating materials. As
was mentioned above, the reason for that is the partial filling of the
$p_{3/2}$ states due to the band degeneracy by the cubic symmetry. On one hand,
a 2D TI can be realized using Heusler compounds by forming QW structures, in which an energy gap
opened due to the confinement effect in the lower dimensions. On the other hand,
a 3D TI can be achieved by introducing strain to the bulk to break the cubic symmetry. This was ever
successfully demonstrated in the bulk HgTe with the in-plane strain~\cite{bruene2011}.

\subsection{Honeycomb compounds}
There is a group of ternary materials that can be view as a honeycomb version of the $XYZ$ Heusler compounds. 
One can make a honeycomb layer formed by $Y$ and $Z$ atoms. Subsequently one can have a stable layered structure 
with a triangle $X$ lattice stuffed in between neighboring $YZ$ honeycomb layers. The low energy bands structures is determined by 
the atomic orbitals of $Y$ and $Z$ atoms, while $X$ is to fulfill the 8- or 18-electron rule. Similar to Heusler compounds, one can design 
a band inversion inside the $YZ$ layer, resulting in a QSH layer. In a 3D lattice of stacked QSH layers, the QSH edge states interact with each other 
on the side surface and form novel topological surfaces states. In such a 3D honeycomb lattice, Yan {\it et al.} predicted the first family of weak TIs~\cite{yan2012prediction},  which have even number of surface Dirac cones on the side surface. 
Since the cubic symmetry is absent, these compounds have a full energy gap.
Among them, KHgSb exhibits a large energy gap of 0.24 eV, providing a system where the topological effects can be observed at the room temperature. If dislocations lines exit in the layered weak TIs, robust line states appears inside the bulk~\cite{ran2009one}, which can be ideal thermoelectric candidates. By fine tuning the SOC strength and layer coupling, strong TIs can also be realized among these honeycomb compounds~\cite{yan2012prediction,ZCM11}, such as LiAuSe.

\begin{figure}[ht]%
  \includegraphics[width=0.7\linewidth]{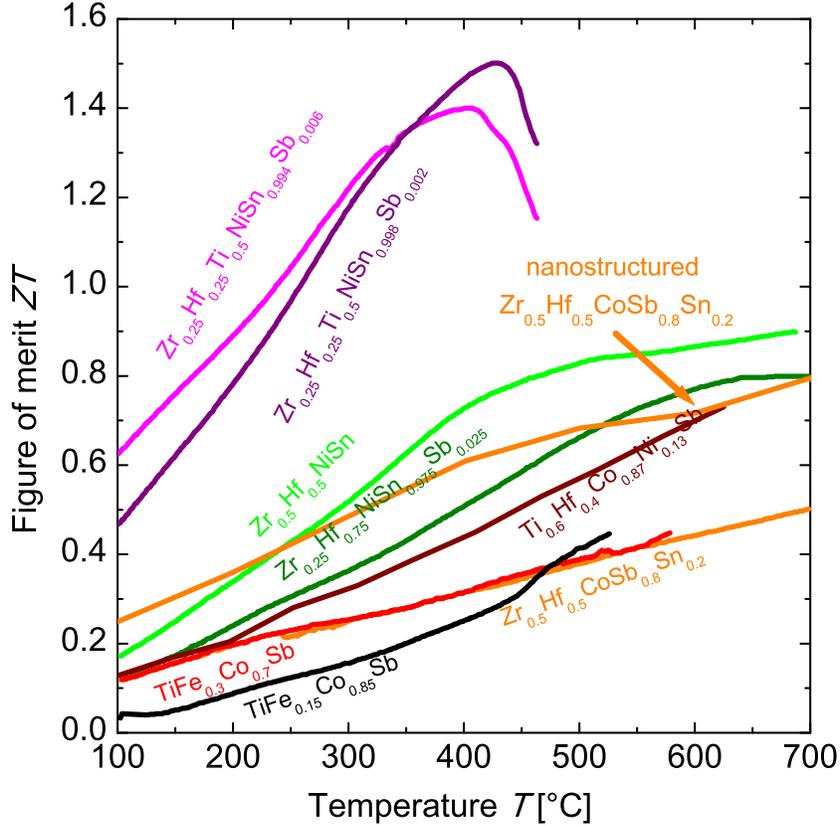}%
  \caption[]{%
    State of the art thermoelectric figure of merit \textit{ZT} of half-Heusler materials.}
    \label{ZTHH}
\end{figure} 

\subsection{Heusler compounds for thermoeletrics}
An advantage of the half-Heusler compounds are the similar chemical and physical properties of the $n$- and $p$-type compounds, making them attractive for thermoelectric modules~\cite{Ro06}.
The research for good thermoelectric compounds within the class of half-Heusler compounds is focused on two systems based on NiTiSn for the $n$-type and
CoTiSb for the $p$-type materials. These compounds can easily be doped with other elements, and thus the band structure can be changed. For NiTi(Sn,Sb) materials, power factors ($S^2\sigma$) up to 70~$\mu$W/cmK$^2$ at 650~K can be reached~\cite{BPL00}. Nevertheless due to the comparatively high thermal conductivity of about 10~$\mu$W/mK, a figure of merit of only 0.45 at 650~K was achieved. One of the highest figures of merit of 1.5 at 700~K was reported for Sb-doped NiTi$_{0.50}$Zr$_{0.25}$Hf$_{0.25}$Sn~\cite{SAK05,SHU05}. The investigation of NiTi$_{1-x}$$M_x$Sn (where $M$ = Sc, V and $0<x\leq0.2$)~\cite{OFB10,OFB10b} and NiTi$_{0.3-x}$Sc$_x$Zr$_{0.35}$Hf$_{0.35}$Sn (where $0<x\leq0.05$) showed a possibility to create $n-$ and $p-$type thermoelectric materials with significantly high power factors and $ZT$s within a single Heusler compound. The $p$-type doping (Sc) creates holes in the triply degenerate valence band at the $\Gamma$-point whereas the $n$-type doping (V) supplies the  electrons to the single conduction band above the indirect gap at the $X$-point. This is typical for all semiconducting transition metal based Heusler compounds in $C1_b$ structure. Further substitutions with other main-group and transition metals have been made for the optimization of the thermoelectric performance~\cite{SOO09,SBa11} (see Fig.~\ref{ZTHH}). 

\subsection{Filled skutterudites}

\begin{figure}[htb]%
  \includegraphics[width=0.7\linewidth]{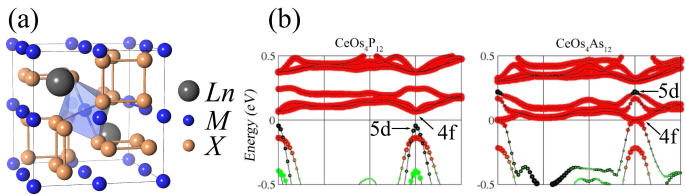}%
  \caption[]{%
    Crystal structure of the filled skutterudites
\textit{Ln}\textit{T}$_4$\textit{X}$_{12}$ (\textit{Ln}= rare-earth, \textit{T}=
Fe, Ru or Os, and \textit{X}=P, As or Sb) (a).
Band structures calculated for CeOs$_4$P$_{12}$~ and CeOs$_4$As$_{12}$~(b).
Red/gray dots stand for the components of Ce
$f$-states, black dots stand for Os $d$-states, and green/light gray stand for
$p$-states. The
size of the dots represents the relative amplitude of corresponding
components.}
    \label{skutterudites}
\end{figure}

Skutterudite is the name of a cobalt and arsenic based mineral that was
extensively mined in the region of Skutterud, Norway.
This compound has the formula CoAs$_3$ and serves as the prototype compound for a whole class of materials. More generally
\textit{MX}$_3$ (\textit{M} = Fe, Co, Rh, Ir, Ni \textit{X}= P, As, Sb) 
compounds with the same cubic crystal  structure have since been known as
'skutterudites'. Fig.~\ref{skutterudites} shows their typical crystal structure
consisting of a bcc lattice of M atoms with \textit{X}$^{4-}_4$ rectangles,
which can be understood using the Zintl concept as explained below:
The transition metal element donates 3 electrons to the three pnictogen atoms
\textit{X}. These \textit{X}$^{-}$ anions are then isoelectronic to oxygen or
sulfur with a \textit{s}$^2$\textit{p}$^4$ configuration.
This means that the  \textit{X}$^{-}$ anions need two more bonds to be in a
\textit{s}$^2$\textit{p}$^6$ closed shell configuration. Hence
\textit{X}$^{4-}_4$ rectangles are formed, which coordinate the metal atoms in
distorted octahedrons. The electronic structure of these materials is then
dominated by the \textit{p}-bands of the \textit{X}$^{4-}_4$ rings and the
\textit{d}-bands of the transition Metal \textit{M}.  A prominent example is
CoSb$_3$ which is a diamagnetic (\textit{d}$^6$ Co$^{3+}$) narrow band gap
semiconductor with ZTs as high as 0.7 \cite{snyder11}.
The skutterudite structure has two voids in the cubic unit cell. Similar to the
Heusler compounds in which the the zinc-blende lattice can be stuffed, filled
skutterudites (FSs) can be obtained by filling the voids with a variety of
atoms, including lanthanides (La, Ce, Pr, Nd, Sm, Eu, Gd, Tb). The FSs have the
chemical formula \textit{Ln}\textit{T}$_4$\textit{X}$_{12}$ (\textit{Ln}=
rare-earth, \textit{T}= Fe, Ru or Os, and \textit{X}=P, As or Sb), in which the
heavy elements are expected to induce strong SOC.
As CoSb$_3$ and Bi$_2$Te$_3$, they are also known for their excellent
thermoelectric properties \cite{sales96} due to the phonon scattering of the
heavy atoms.
Because of the rare earth atoms, their electronic structure is immensely
complex, ranging from simple hybridization gap semiconductors to  a rich variety
of electronic and magnetic ground states at low temperature, including
superconductivity, ferromagnetism, and Kondo insulating behavior
\cite{Sal+03,MHB+08,SSA+09}.
The Ce-based compounds CeOs$_4$\textit{X}$_{12}$ are reported to be
semiconductors due to the hybridization of the Os-\textit{d} and Ce-\textit{f}
states. It turns out, that the P containing compound is a simple hybridization
gap semiconductor, whereas the As and Sb containing compounds are semimetals
with an inverted band structure \cite{YMQ+11}. In these compounds, the
\textit{d}- and \textit{f}-bands invert at the $\Gamma$-point, which results in
an semimetal due to cubic symmetry, similar to HgTe and the Heusler compounds
(See Fig.~\ref{skutterudites}). Again, there are two ways to turn them into
bulk-insulating TIs. On the one hand, it should be possible to fabricate quantum
well structures from CeOs$_4$P$_{12}$ and CeOs$_4$As$_{12}$ or
CeOs$_4$Sb$_{12}$, which will result in a 2D TI. On the other hand, the
degeneracy can be lifted by strain while maintaining the BI. 
So far, BIs have only been observed between \textit{s}- and \textit{p}-bands
(HgTe, Heuslers) or \textit{p}-bands of different parity. In the family of
filled skutterudites however, the \textit{d}- and \textit{f}-bands invert.
At low temperatures $T<135$~K for CeOs$_4$As$_{12}$and $T<50$~K for
CeOs$_4$Sb$_{12}$)~\cite{LKa+98,KWL+11}, the inverted compounds have also been
reported to be Kondo insulators. If the band inversion remains stable, these
materials could realize a new topological nontrivial phase called topological
Kondo insulators, where the surface states would exist within the Kondo-gap
\cite{CQK+10,LWX+10}.

\section{The Bi$_2$Se$_3$ family of compounds}

\begin{figure}[ht]%
  \includegraphics[width=0.7\linewidth]{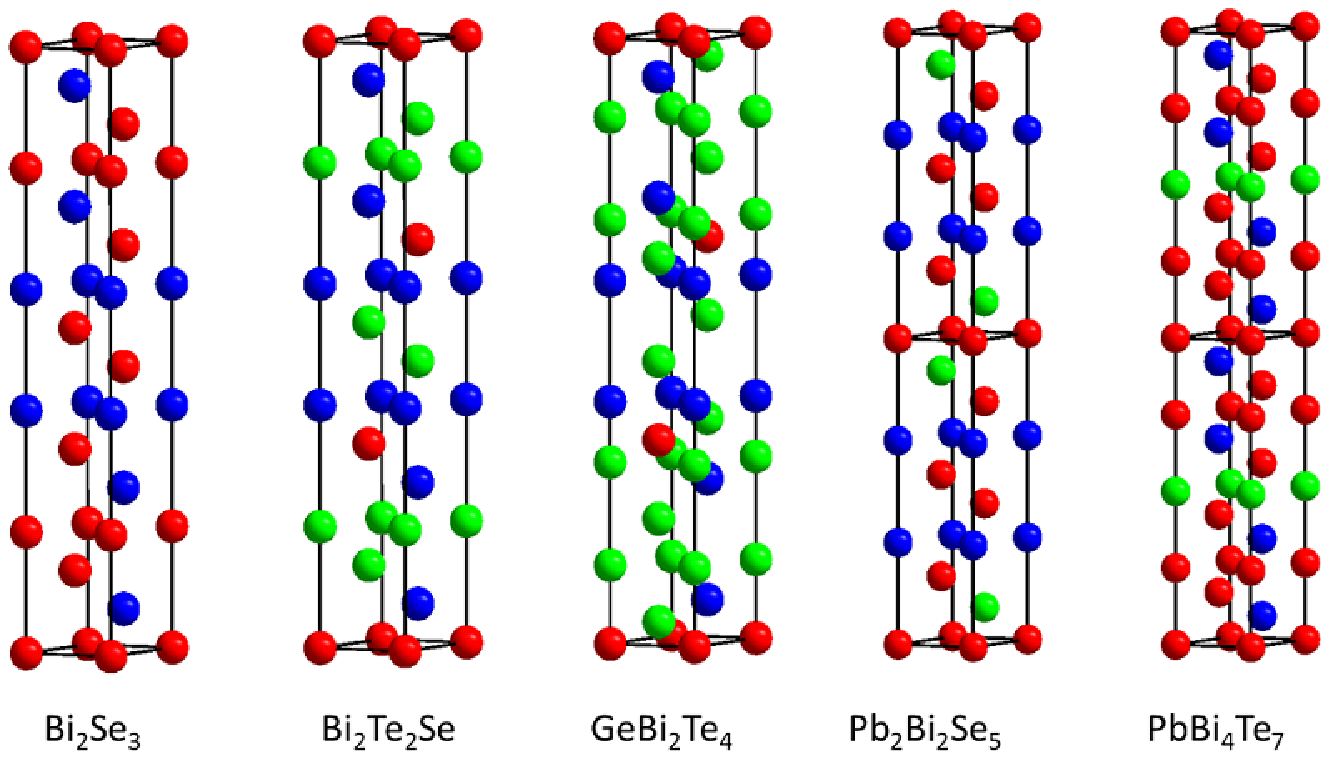}%
  \caption[]{%
    Crystal structure of Bi$_2$Se$_3$, Bi$_2$Te$_2$Se, GeBi$_2$Te$_4$, Pb$_2$Bi$_2$Se$_5$ and PbBi$_4$Te$_7$ }
    \label{HexStructures}
\end{figure}

The first 3D TI discovered was Sb doped Bi \cite{Hsieh08}.Shortly after this
discovery however, Bi$_2$Se$_3$, Bi$_2$Te$_3$ and Sb$_2$Te$_3$ were found to be
TIs with a much simpler electronic structure \cite{ZLQ+09,XQH+09,CAC+09}.
Today the Bi-Te-Sb family is considered to be the prototype family of 3D TIs.
Many compounds with related structures have been proposed and identified as 3D
TIs, however most of the attention is focused the Bi$_2$Se$_3$ class of
compounds. 
The layered semiconductors, TlBiTe$_2$ and TlBiSe$_2$, were first theoretically predicted~\cite{YZL+10} to be TIs and soon observed in experiments~\cite{SSG+10,KYK+10,CLA+10}.
Very recently a systematic study of related structures such as LaBiTe$_3$~\cite{yan2010a}, Bi$_2$Te$_2$Se~\cite{ren2010}, GeBi$_2$Te$_4$~\cite{NXW12}, Pb$_2$Bi$_2$Se$_5$, and PbBi$_4$Te$_7$~\cite{NXW12,eremeev2010,jin2011} have been reported to feature multifunctional and flexible electronic structures. 
These compounds are all structurally related to Bi$_2$Se$_3$ (Fig.~\ref{HexStructures}). Bi$_2$Se$_2$Te, Bi$_2$Te$_2$Se, Bi$_2$Te$_{1.5}$S$_{1.5}$, and GeBi$_2$Te$_4$ exhibit single Dirac cone surface states, which is supported by first principles band calculations.
Among them, Bi$_2$Se$_2$Te, Bi$_2$Te$_{1.5}$S$_{1.5}$, and GeBi$_2$Te$_4$ feature an in-gap Dirac point~\cite{NXW12}. 

\begin{figure*}[htb]%
  \includegraphics[width=0.7\linewidth]{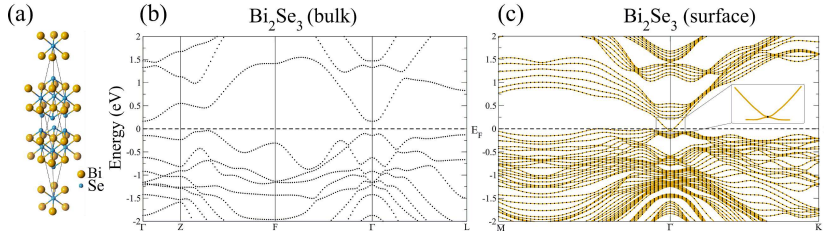}%
  \caption[]{%
    Crystal structure of Bi$_2$Se$_3$ (a). Bulk band structure
with SOC included (b). Band structure of 5 quintuple layers of Bi$_2$Se$_3$ with
a single Dirac cone (c). The inset shows the Dirac cone and the Dirac point,
which is the point of the Kramer's degeneracy, indicated by a black dot}
    \label{Bi2Se3}
\end{figure*}

Bi$_2$Te$_3$ has been known to be an excellent thermoelectric compound for quite
some time and is routinely used in devices.
The crystal structure consists of quintuple layers composed of Bi and the
chalcogenide which are layers of edge-sharing SeBi$_6$ octahedra (see
Fig.~\ref{Bi2Se3} (a)). All compounds within this family are small band gap
semiconductors due small differences in electronegativity and SOC. Because of
the heavy atoms, phonon-scattering and relativistic effects are greatly enhanced
in this family favoring both thermoelectric and topological properties. The
small band gaps of the order of 0.3 eV result in higher Seebeck coefficients and
good power factors and allow the study of the surface states at elevated
temperatures.
In this class of compounds, the inversion occurs between the the
Bi-\textit{p}$_z$ and chalcogenide Bi-\textit{p}$_z$ bands of different parity,
resulting in a single Dirac cone on the surface \cite{ZLQ+09}.
In contrast to compounds used in thermoelectrics, single crystals with control
over the charge carrier density are needed to study the topological effects. In
Bi$_2$Se$_3$ and Bi$_2$Te$_3$ however, the defect chemistry is rather hard to
control: Bi$_2$Se$_3$ is \textit{n}-doped due to charged Se vacancies, whereas
Bi$_2$Te$_3$ can be \textit{n}- or \textit{p}-doped due to antiside defects \cite{Jia11} and
therefore the undoped samples naturally show metallic transport properties.
Recently samples of Bi$_2$Te$_2$Se
\cite{Jia11,TSA+10}, (Bi$_{1-x}$Sb$_x$)$_2$Te$_3$~\cite{KCC+11,ZCZ+11} or
Bi$_2$Te$_{1.6}$S$_{1.4}$ \cite{Huiwen12} have been grown, which exhibits better bulk-insulating properties, making it possible to study the transport of surface states without the influences of the bulk.

\section{Summary}

Some of the key ingredients for both thermoelectrics and topological insulators are the same. A low thermal conductivity and a high electrical conductivity are the main factors of a excellent thermoelectric material. A higher electrical conductivity can be achieved by narrow band-gap materials, while the thermal conductivity can be reduced by atoms with a large atomic mass. Topological insulators must have strong spin orbit coupling, which is increasing with the atomic mass. The inverted band structure comes together with a narrow band gap. Therefore it is not a coincidence that most topological insulators are also good thermoelectric materials. All topological insulators exhibit excellent thermoelectric properties, but on the other hand many thermoelectrics are topologically trivial. For example, PbTe is  known to be a good thermoelectric material, however it is not a topological insulator~\cite{fu2007}. 
The band inversion of the \textit{s}- and \textit{p}-states takes place on the $L$ point, which appears four times in the reciprocal space, so there is an even number of BIs.
Although the understanding of the direct theory behind this relationship of topological insulators and thermoelectric materials is incomplete, the correlation between them should guide the discovery of new topological insulators and thermoelectrics. On the other hand, the topological states~\cite{Takahashi2010,ghaemi2010}  may be a new quantity to optimize the figure of merit $ZT$ and realize high performance thermoelectric devices.

%
%



\end{document}